\documentclass[12pt,preprint]{aastex}

\begin{document} 

\title{
The G11.11-0.12 Infrared-Dark Cloud: 
Anomalous Dust and a Non-Magnetic Isothermal Model
}

\author{Doug Johnstone\altaffilmark{1}, 
Jason D. Fiege,\altaffilmark{1},  
R.O.\ Redman,\altaffilmark{1},  
P.A.\ Feldman,\altaffilmark{1}, 
and Sean J. Carey\altaffilmark{2}} 

\altaffiltext{1}
{National Research Council Canada, Herzberg Institute of Astrophysics,
5071 West Saanich Rd, Victoria, BC, V9E 2E7, Canada;
doug.johnstone@nrc-crnc.gc.ca,
jason.fiege@nrc-crnc.gc.ca,
russell.redman@nrc-crnc.gc.ca,
paul.feldman@nrc-crnc.gc.ca
}

\altaffiltext{2}
{SIRTF Science Center, California Institute of Technology, Mail Code 
220-6, Pasadena, CA 91125; carey@ipac.caltech.edu}

\begin{abstract}
The G11.11-0.12 Infrared-Dark Cloud has a filamentary appearance, both 
in absorption against the diffuse 8$\mu$m Galactic background, and in 
emission from cold dust at 850$\mu$m. Detailed comparison 
of the dust properties at these two wavelengths reveals that standard
models for the diffuse interstellar dust in the Galaxy are not
consistent with the observations. The ratio of 
absorption coefficients within the cloud is 
$\kappa_8/\kappa_{850} \le 1010$, which is well below 
that expected for the diffuse ISM where $\kappa_8/\kappa_{850} \sim 1700$.  
This may be due to the formation of ice mantles on the dust and grain 
coagulation, both of which are expected within dense regions of molecular 
clouds.  The 850$\mu$m emission probes the underlying radial structure 
of the filament.  The profile is
well represented by a marginally resolved central region 
and a steeply falling envelope, with $\Sigma(r) \propto
r^{-\alpha}$, where $\alpha \geq 3$, indicating that G11.11-0.12 is the 
first observed filament with a profile similar to that of a non-magnetic 
isothermal cylinder.  
\end{abstract}


\section{Introduction and Observations}

Infrared-dark clouds (IRDCs) have been 
identified by their substantial mid-infrared (8-25$\mu$m) extinction
in {\it Midcourse Space Experiment} (MSX) 
images \citep{ega98}. Egan et al.\ concluded  from the
high mid-infrared opacities  of the IRDCs that they possess hundreds
of magnitudes of visual extinction and  contain large column densities
of cold dust. They derived characteristic physical parameters 
for the gas inside the IRDCs: $l\,\approx$~0.5 -- 15\,pc, 
$T_{\rm K}$~$\approx$~10 -- 20\,K, n(H$_2$)~$\ge$~10$^5$~cm$^{-3}$, 
and H$_2$ column densities ranging up to 10$^{23}$~cm$^{-2}$ \citep{car98}.  

A number of IRDCs have been mapped with SCUBA on the JCMT \citep{car00, 
red00, red03} and are found to have the mid-infrared absorption closely 
associated with filamentary and/or floculent clouds in emission at 850 and
450$\mu$m. Bright, compact sources with masses in the range 
$\approx$~10 to 10$^3$~M$_\odot$ and infall/outflow signatures are
located along the  filaments and appear to be in a variety of the early
stages of star formation.  The larger IRDCs appear to be similar in size
to the Orion molecular cloud but different in many important 
aspects. They are more heavily extincted, colder, more quiescent, and at 
earlier stages of evolution.

\begin{figure}[!htp]
\plotone{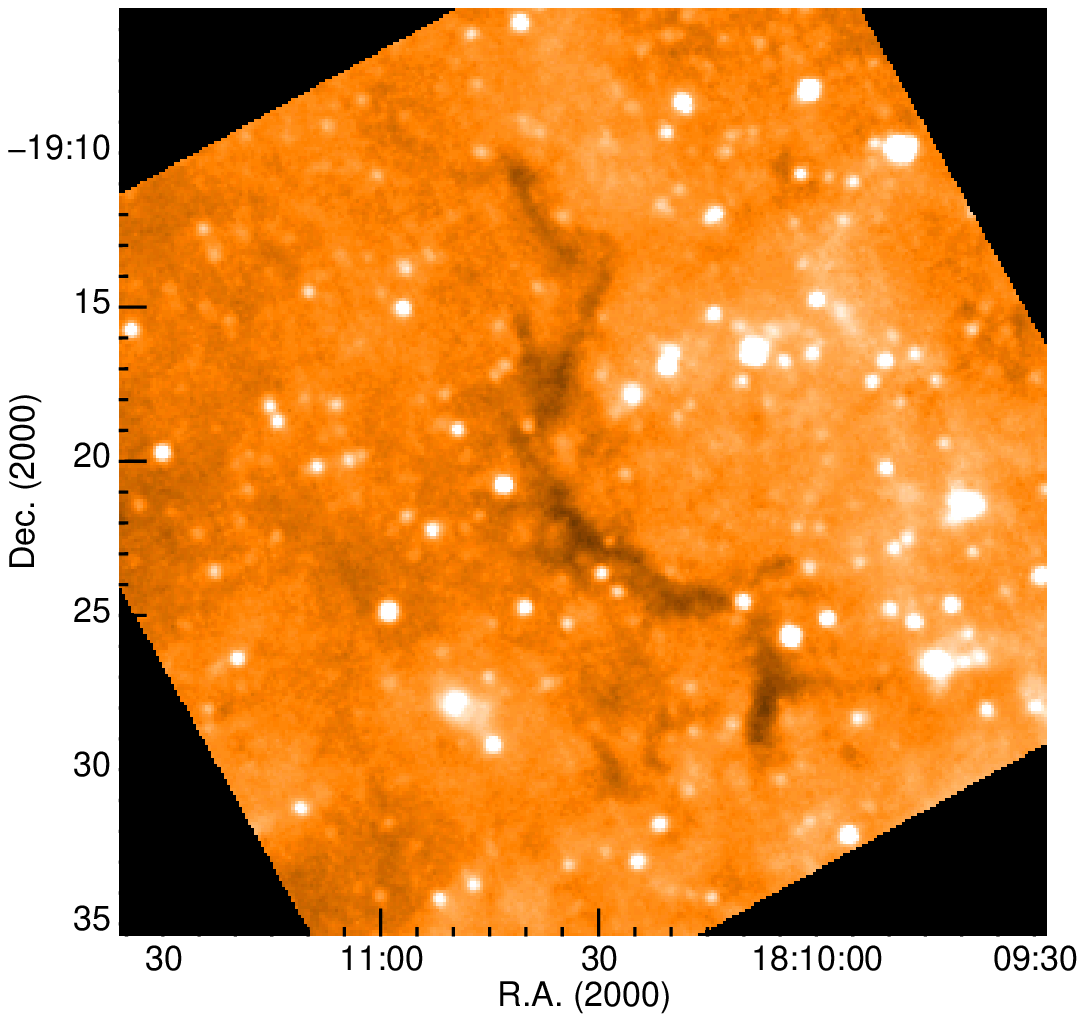}
\caption{MSX 8$\mu$m image showing the mid-infrared emission in the 
direction of G11.11.  Note the narrow filamentary extinction lane. 
\label{f_8}}
\end{figure}

\begin{figure}[!htp]
\plotone{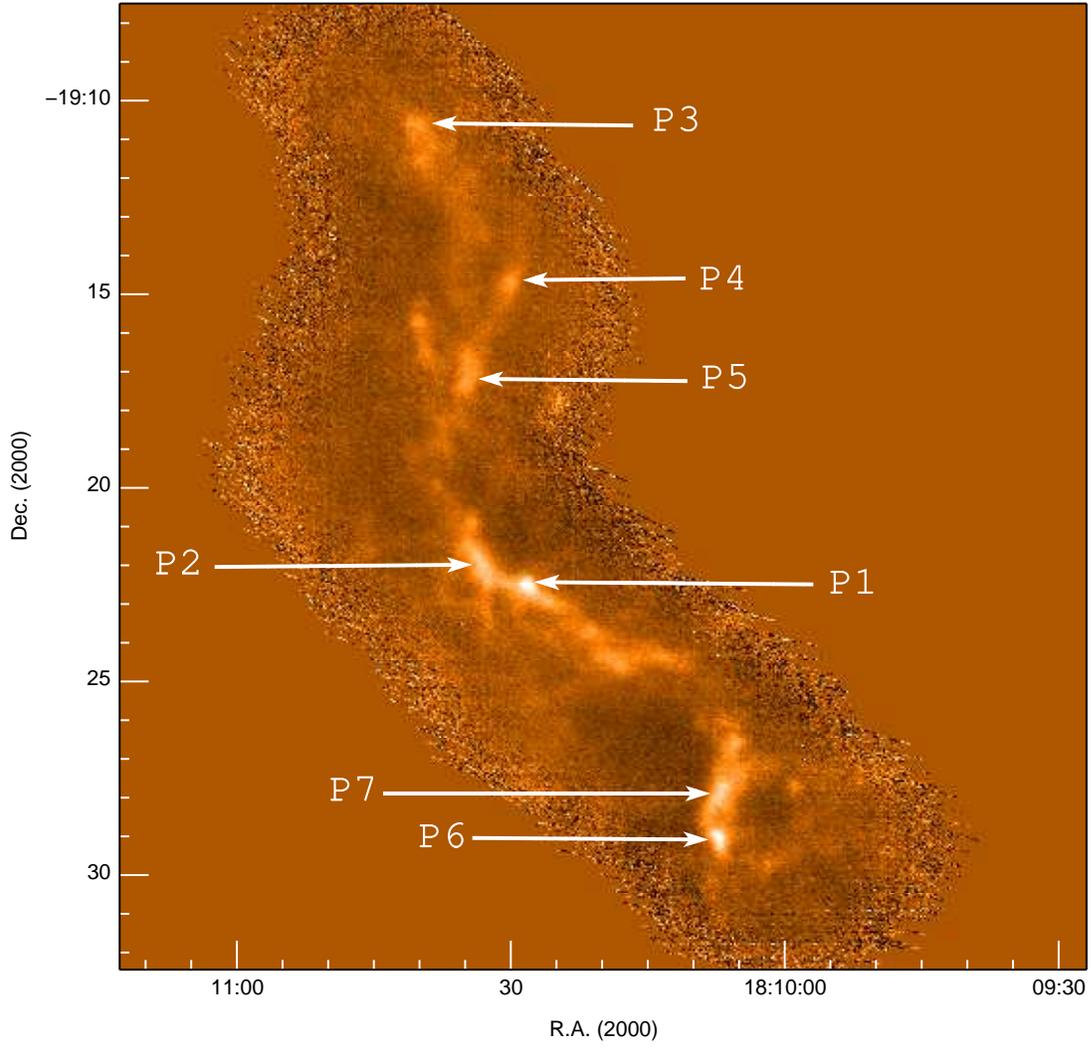}
\caption{850$\mu$m emission in the direction of G11.11.  The intensity range 
is from -0.1 to 0.7 Jy/beam.  Note the strong correlation between the 
submillimeter dust emission and the mid-infrared extinction in 
Fig.\ \ref{f_8}.  
\label{f_850}}
\end{figure}

For the work reported in this paper, we investigate the IRDC G11.11-0.12,
at a kinematic distance of 3.6~kpc \citep{car98}, because it contains
several long, filamentary segments seen against a relatively uniform
galactic background at 8$\mu$m (Fig.\ \ref{f_8}). 
\citet{car00} first reported JCMT observations of this source.
Star formation appears to be more active in the southern
part of the  cloud than the northern half, with bright, compact SCUBA
sources (labelled P1, P6 and P7 in Fig.\ \ref{f_850}). The P1 source is driving 
a molecular outflow and is associated with a weak point source of 8$\mu$m
emission.  There is no evidence of high-mass star formation throughout most 
of the cloud.  The catalog of \citet{bec94} lists two UCHII
regions in the vicinity of G11.11-0.12, but they are far enough away that they
are unlikely to be affecting the IRDC properties even if they are at the same
distance.

JCMT\footnote{The JCMT is operated by the Joint Astronomy Centre on behalf 
of the Particle Physics and Astronomy Research Council of the UK, the 
Netherlands Organization for Scientific Research, and the National Research 
Council of Canada} SCUBA observations were obtained at 
850$\mu$m under fair weather conditions
($\tau_{850}$= 0.24 -- 0.30) in April 1999 with an angular
resolution of 14.5$^{\prime\prime}$.  ``Scan mapping'' was employed to
produce  Nyquist-sampled images. The data were taken with the standard
observing set-up using chop throws of 20$^{\prime\prime}$,
30$^{\prime\prime}$  and 65$^{\prime\prime}$ at position angles of 0$^o$
and 90$^o$. The data were processed using the JCMT data reduction pipeline 
(ORAC) to flatfield, extinction correct and calibrate the chop data 
\citep{ebw99, je99}.  The calibration accuracy is good to $\pm 15$\% 
\citep{jsa02}.  The optimal map reconstruction (see Fig.\ \ref{f_850})
was then performed from the 
individual chop measurements, using a matrix inversion technique outlined 
by \citet{jmi00}.
Slowly varying spatial features with wavelengths much longer than the maximum
chop,  65$^{\prime\prime}$, are not measured reliably; thus, large, 
low-amplitude structures in the reconstructed map tend to be artifacts.
Further descriptions of SCUBA and its observing modes can  be found 
in \citet{hol99} and \citet{jen00}.

Heterodyne observations of $^{13}$CO~($2-1$) and C$^{18}$O~($2-1$) 
were taken near the P2 region of G11.11-0.12 with receiver RxA3 on the JCMT 
during several observing sessions in July 2000,  April 2001, 
and June 2001.  The velocity dispersion within the cloud can be 
estimated by fitting single component Gaussians to the line profiles.  
Typical linewidths obtained have $\sigma = 1.2\,$km$\,$s$^{-1}$ for 
the $^{13}$CO line and $\sigma = 0.9\,$km$\,$s$^{-1}$ for the C$^{18}$O line.
The temperature of the gas in the segment of the filament labelled P2 
(Fig.\ \ref{f_850}) has been estimated from H$_2$CO observations by
\citet{car98} to be less than 20\ K.

\section{Dust Properties: Extinction versus Emission}

The extinction of background emission at 8$\mu$m is due to the presence of dust 
within the G11.11-0.12 filament. This same dust radiates at 
submillimeter wavelengths and is observable as emission in the 
850$\mu$m image. It is expected that there
should be a strong correlation between the extinction and emission maps, and 
this is indeed seen in Figs.\ \ref{f_8} and \ref{f_850}. The 
correlation is complicated by several factors: (1) the diffuse background 
emission at 8$\mu$m, which the molecular cloud extinguishes, is not 
entirely smooth; (2) star formation occurring at P1 produces 8$\mu$m emission; 
(3) the zero submillimeter emission baseline is poorly defined; (4) the 
measured emission and absorption properties are beam-convolved; (5) the dust 
temperature and emissivity may change with location in the
cloud because the degree of dust coagulation and freeze-out of 
molecules onto grain mantles depend on the local physical conditions, such as 
density and temperature.

\begin{figure}[!htp]
\epsscale{1.0}
\plotone{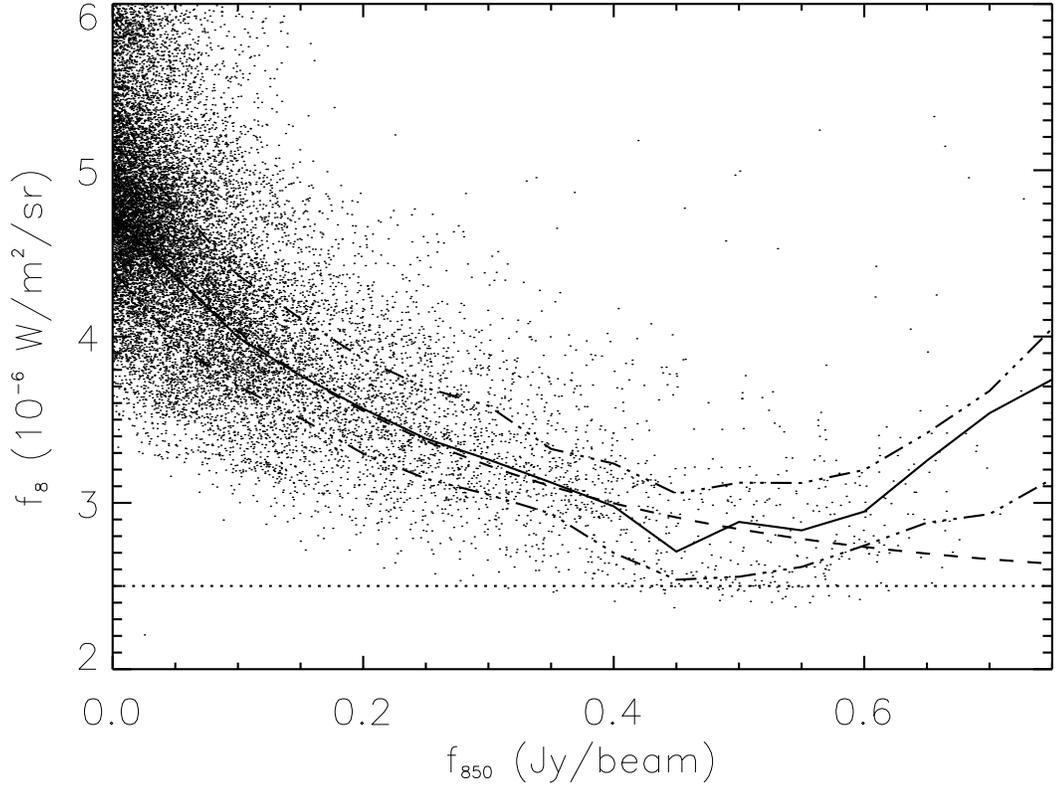}
\caption{Measured 8$\mu$m flux versus 850$\mu$m flux at each position within 
the G11.11 filament. The solid line indicates the median 8$\mu$m flux as a 
function of 850$\mu$m flux while the dashed-dotted lines give the range 
containing 50 percent of the 8$\mu$m measurements. The overlaid dashed line 
represents the best-fit model (see text) with the dotted line marking the
required 8$\mu$m foreground flux 
$f_{8}({\rm fg})= 2.5\times 10^{-6}$\,W\,m$^{-2}$\,sr$^{-1}$. Note that 
the fit becomes poor
at large 850$\mu$m flux due to the presence of a warm source at P1, producing
internal 8$\mu$m emission.
\label{f_dust}} 
\end{figure}

Statistically relevant results can be obtained by analyzing a large 
sample of measurements, despite the above caveats.  Fig.\ \ref{f_dust}
shows the 8$\mu$m emission versus the 850$\mu$m emission for each location 
across the filament, along with the median 8$\mu$m flux as a function of 
850$\mu$m emission. The trend in both the data and the median curve is clear:
the 8$\mu$m flux decreases due to extinction as the 850$\mu$m flux increases.
The highest 850$\mu$m emission points do not follow this trend
because of the presence of an energetic protostellar source (P1) within the deepest
part of the cloud that is altering the temperature of the
dust and producing weak 8$\mu$m emission. 

Given the clear trends in Fig.\ \ref{f_dust}, 
a simple dust model may be fit to the 
data.  At each location ${\bf x}$ across the filament, the measured 
8$\mu$m flux density $f_8({\bf x})$ 
depends on an unvarying foreground $f_{8}({\rm fg})$ and background 
$f_{8}({\rm bg})$ emission, and the optical depth through the cloud 
$\tau_8({\bf x}) = \kappa_8\,\Sigma({\bf x})$, where 
$\kappa_\lambda$ is the absorption coefficient of the gas plus dust at 
wavelength $\lambda$($\mu$m) and $\Sigma({\bf x})$ is the column density of 
material within the intervening cloud. Explicitly,
\begin{equation}
f_8({\bf x}) = f_{8}({\rm bg})\exp\left[-\kappa_8\,\Sigma({\bf x})\right] 
+ f_{8}({\rm fg}).
\end{equation}
The column density 
of the cloud is related to the the observed flux density $f_{850}({\bf x})$
(measured in Jy/beam), for the optically thin submillimeter emission, 
by the equation
\begin{equation}
f_{850}({\bf x}) = 
B_{850}(T_d)\,\kappa_{850}\,\Sigma({\bf x})\,\Omega,
\end{equation}
where $B_\lambda(T_d)$ is the Planck function at $\lambda$ and 
$\Omega$ is the solid angle subtended by the telescope beam.
Thus, the optical depth at 8$\mu$m, which is given by 
$\tau_8({\bf x}) = k\,f_{850}({\bf x})$,
is linearly related to the observed flux at 850$\mu$m, with $k$
defined by
$\kappa_8/\kappa_{850} =  k\,B_{850}(T_d)\, \Omega$.

The correlation between the flux at 8$\mu$m and 850$\mu$m in 
Fig.\ \ref{f_dust} allows for a determination of $k$ 
(and $f_{8}({\rm bg})$, $f_{8}({\rm fg})$).  The overlaid dashed line in 
Fig.\ \ref{f_dust} is based on the best-fit model, with $k = 3.75$, 
$f_{8}({\rm bg})= 2.3\times 10^{-6}$\,W\,m$^{-2}$\,sr$^{-1}$, and 
$f_{8}({\rm fg})= 2.5\times 10^{-6}$\,W\,m$^{-2}$\,sr$^{-1}$. 
Excellent fits are found for a small range around $k = 3.75\,\pm0.25 $, 
implying 
\begin{equation}
{\kappa_8 \over \kappa_{850}} \simeq  { 1350 \over \exp(17\,{\rm K}/T_d) - 1}.
\end{equation}
Observations of molecular lines provide a range of gas temperatures,
$T_{\rm K}$~$\approx$~10 -- 20\,K, with a best-fit value of $T_{\rm K} = 15\,$K.
Theoretical models of the dust temperature toward highly extincted cloud 
centers \citep{zwg01, ers01, krm03} yield lower limits of 5 -- 8\,K. 
The corresponding limits on the opacity ratio are 1010 if $T_d = 20\,$K and 
50 if $T_d = 5\,$K.

For diffuse interstellar dust the expected opacity ratio is
$\sim 1700$ \citep{ld01}; however, ice mantles
form on the surfaces of dust grains in dense molecular clouds 
and the grains themselves coagulate. 
This increases the submillimeter opacity and obscures the silicate features 
that account for most of the opacity at 8$\mu$m \citep{oh94}. Ossenkopf
and Henning found that, after $10^5\,$yrs of coagulation, dust with thin 
ice mantles entrained in cold ($T_{\rm K} < 20\,$K)  gas with number density
$n = 10^6\,$cm$^{-3}$ produced an opacity ratio between 
8$\mu$m and 850$\mu$m of $\sim 500$, consistent with these observations
if the dust temperature is $T_d \simeq 13\,$K. 
\citet{krm03} found a similar increase in the importance
of submillimeter emission compared to (near) infrared extinction
in the filamentary cloud IC 5148. These result are consistent with
evidence for dust growth in cold, dense protoplanetary disks
where observations of the slope of the spectral energy distribution are
used \citep{ll03}

All reasonable dust  models require a foreground 8$\mu$m emission
of $f_8({\rm fg}) \sim f_8({\rm bg})$ which implies that half of the diffuse
mid-infrared emission is produced {\it in front} of the filament. This
is intriguing given that G11.11-0.12 is only 3.6 kpc away and on the near side 
of the Galactic Center. The diffuse 8$\mu$m emission, while relatively 
uniform across the IRDC, must be significantly patchy on Galactic scales.

\section{The Filamentary Structure of G11.11-0.12}

The simplest equilibrium model for a filament is that of an isothermal cylinder 
\citep{o64} where the density profile as a function of cylindrical radius $r$
is given by 
\begin{equation}
\rho = {\rho_c \over \left(1 + r^2/8r_0^2\right)^2},
\end{equation}
where $\rho_c$ is the central density and $r_0$ is the core radius defined by
$\sigma^2 = 4\,\pi\,G\,\rho_c\,r_0^2$.
The velocity dispersion $\sigma$ should be derived from the temperature of 
the gas for a true isothermal cylinder. However, in most astrophysical cases 
the measured velocity dispersion within the molecular cloud includes a 
significant ``non-thermal" (turbulent) component. 

The surface density profile for an edge on isothermal cylinder is:
\begin{equation}
\Sigma = {\Sigma_0 \over \left(1 + r^2/8r_0^2\right)^{3/2}},
\end{equation}
where $\Sigma_0 = 2^{1/2}\,\pi\,\rho_c\,r_0$. 
The surface density profile becomes $\Sigma \propto r^{-3}$,
at large projected distances from the center of the cylinder.

Most studied filamentary clouds are not well-fit by this
profile.  L97 in Cygnus \citep{all98}, the Integral Shaped Filament (ISF) in 
Orion A \citep{jb99}, IC5146 \citep{lal99}, and a quiescent filament in 
Taurus \citep{sab03} are all better fit by shallow density profiles with 
$\rho \propto r^{-2}$, which are consistent with surface density profiles of 
$\Sigma \propto r^{-1}$. Such profiles are well-fit by the \citet{fp00} model, 
which features a helical magnetic field. The $r^{-2}$ density profile is 
obtained in FP models with a dominant toroidal field
component, but steeper profiles are possible when the poloidal field 
dominates. Submillimeter polarization measurements trace the geometry of the
magnetic field and provide additional evidence in support of the FP helical 
field model, which has been shown to agree with polarization maps of the ISF
\citep{mwf01} and NGC2024 \citep{mfm02}.

\begin{figure}[!htp]
\epsscale{1.0}
\plotone{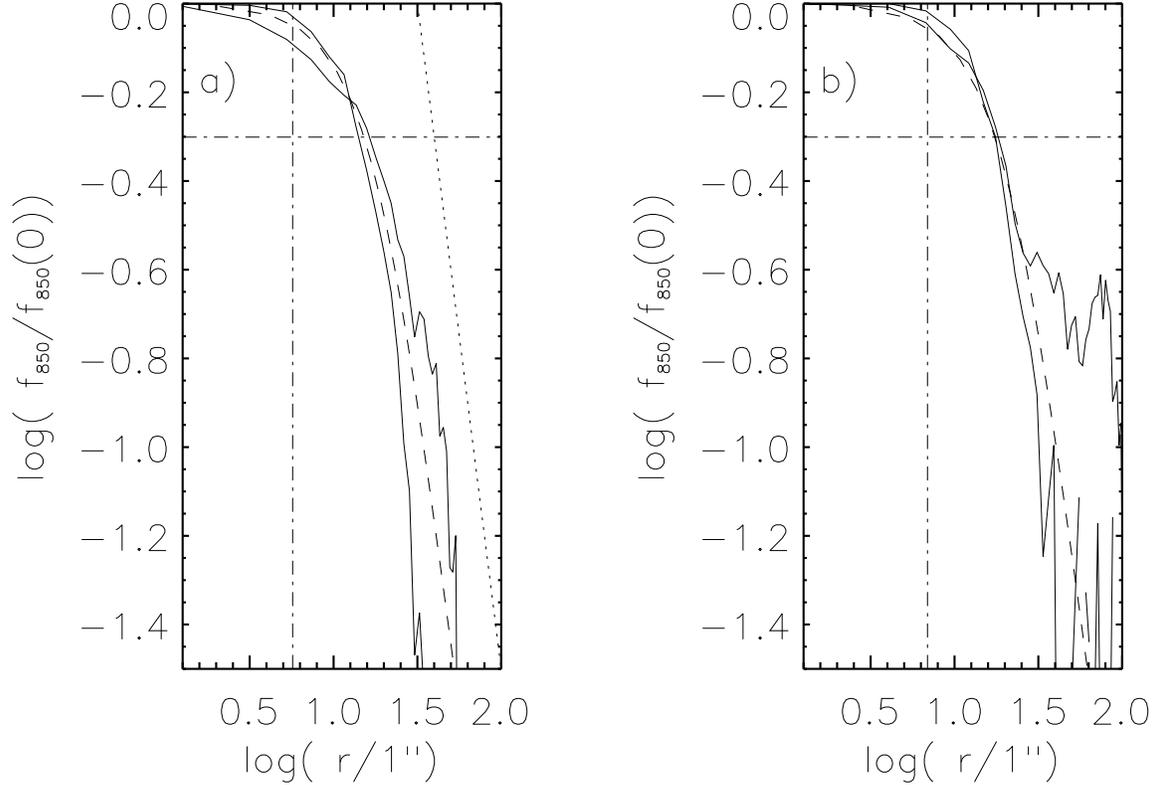}
\caption{850$\mu$m emission profile across the G11.11 filament measured at 
two locations: (a) southern segment, (b) northern segment.  In each panel, the 
two solid lines show independently the left and right profile from the ridge 
center. The overlaid curved dashed line represents the column density profile 
of an isothermal cylinder with $r_0 = 0.10\,$pc (southern segment) and 
$r_0 = 0.12\,$pc (northern segment) after convolution by the $14.5''$ beam of 
the JCMT. The dash-dotted lines denote where the flux is half the peak value 
and the radius corresponding to the $14.5''$ beam.  The dotted line in 
panel (a) shows a power-law relation with $f \propto r^{-3}$, equivalent to
the asymptotic isothermal profile. In panel (b), the flattening of
one profile is due to the proximity of a second filamentary spur.
\label{f_profile}}
\end{figure}

The radial structure of two well-defined regions of the
G11.11-0.12 filament, a northern segment including P3 and P4, and a southern
segment including P1 and P2, have been analyzed.  The average intensity
distributions orthogonal to the spine of the filament were determined
using the same technique as applied to the ISF \citep{jb99}, and are 
presented in Fig.\ \ref{f_profile}. 
The G11.11-0.12 segments appear similar to each other, 
with a marginally resolved central region connecting to an outer region
characterized by a density profile steeper than $r^{-3}$. This
is in contrast to results for the ISF where the central region is
unresolved, and whose surface density only falls off as 
$r^{-1.0\pm 0.25}$ at large radii $r \gg r_0$
\citep{jb99}.  

The emission from each section of G11.11-0.12 is reasonably fit 
by an isothermal profile (dashed lines in Fig.\ \ref{f_profile}) 
allowing for an estimation of the core radius $r_0$.
The core radius varies from $r_0 = 0.12\,\pm\,0.02\,$pc in the northern
segment to $r_0 = 0.10\,\pm\,0.02\,$pc in the southern segment (assuming 
a $3.6\,$kpc distance to G11.11-0.12). The measured peak flux is 
$f_0 = 0.20\,\pm\,0.03\,$Jy/beam and $f_0 = 0.43\,\pm\,0.06\,$Jy/beam, 
respectively, where the uncertainty in the peak flux is primarily due to
uncertainties in the 850$\mu$m calibration. The observed 
dust emission can be converted to a central surface density $\Sigma_0$, 
despite uncertainty in both the dust temperature and emissivity, using  
\begin{equation}
\label{e-sigma}
\Sigma_0 = 0.13\,
	  \left( {f_0 \over 1\, {\rm Jy/beam}}  \right)\,
          \left( \kappa_{850} \over 0.02 \,{\rm cm}^2\,{\rm g}^{-1} 
	  \right)^{-1}\,
	  \left[ \exp(17\,{\rm K}/T_d) - 1 \right]\ \ {\rm g\,cm}^{-2}.
\end{equation}
Assuming a typical dust temperature of $13\,$K, a gas-to-dust ratio of 100, 
and a dust emissivity $\kappa_{850} = 0.02 \,{\rm cm}^2\,{\rm g}^{-1}$ 
\citep{oh94}, the central surface density is $0.07\,\pm\,0.01\,$g\,cm$^{-2}$ 
in the northern segment and
$0.15\,\pm\,0.02\,$g\,cm$^{-2}$ in the southern segment.  If the dust 
temperature were much lower, the column density would be significantly larger. 

The central number density of H$_2$ atoms can also be derived from the 
profile:
\begin{equation}
n_c = 2.6 \times 10^{4}\,
	  \left( {f_0 \over 1\, {\rm Jy/beam}}  \right)\,
	  \left( {r_0 \over 0.1\, {\rm pc}} \right)^{-1}\,
          \left( \kappa_{850} \over 0.02 \,{\rm cm}^2\,{\rm g}^{-1} 
	  \right)^{-1}\,
	  \left[ \exp(17\,{\rm K}/T_d) - 1 \right] {\rm cm}^{-3}.
\end{equation}
For the canonical dust properties, $n_c = 1.1\,\pm\,0.3\times 10^4\,$cm$^{-3}$ 
in the northern segment, and $n_c = 3.3\,\pm\,0.7\times 10^4\,$cm$^{-3}$ in 
the southern segment.  The measured C$^{18}$O linewidth observed for G11.11-0.12 
near the location of P2 is $\sigma = 0.9\,$km/s, allowing for an 
independent measure of the central density, 
$n_c = 3.4\,\pm0.6\times 10^4\,$cm$^{-3}$, via the core radius definition.
The agreement between these two derivations of central density provides
support for the isothermal model. However, more complex filamentary 
models, including poloidal and toroidal magnetic components, may produce 
as good or better fits to the data \citep{fj03}.

\section{Discussion and Conclusions}

G11.11-0.12 is the first molecular filament observed to have a radial 
density profile steeper than $r^{-3}$ at large radii $r \gg r_0$.
A non-magnetic isothermal cylinder model \citep{o64} provides a 
reasonable fit to both the profile and the
measured constraints: submillimeter flux, temperature, velocity dispersion, 
and core radius.  The data might also be consistent with
the generalized magnetic Stodolkiewicz model \citep{s63,nhn93},
whose density profile is identical to that of the Ostriker model, except that 
the core radius definition takes into account the
effects of a helical magnetic field.  The data are not consistent with the
toroidally dominated regime of the FP model, which results in much more
shallow density profiles \citep{fp00}.  Work is in progress to provide a
detailed parameter space exploration and numerical fits to the Ostriker model,
the generalized Stodolkiewicz model and the full range of the FP model,
including the steep profiled regime that is dominated by the poloidal
field component.

The IRDCs are thought to be a cold, dense, and quiescent class of molecular 
clouds \citep{ega98, car98, car00}. The dust properties measured for G11.11-0.12 
between the mid-infrared 
and the submillimeter are consistent with this interpretation; however, 
the theoretical models of \citet{oh94} require approximately 10$^5\,$yrs
for grain coagulation to become important and thus place a lower
limit on the  lifetime of the cloud. The dynamical time of a self-gravitating
non-magnetic isothermal filament, $t_d \simeq (G\rho_c)^{-1/2}$, on which the filament might
be expected to fragment and form stars, is similar to the crossing time 
$t_c \simeq 2\,r_0/\sigma$, which can be calculated directly from the 
observations; $t_c \simeq 2 \times 10^5\,$yrs.  Thus, the required 
coagulation time does not pose severe constraints on the equilibrium
of the filament.  It is harder to reconcile the filament's moderate central 
density $n_c = 1-4 \times 10^4\,$cm$^{-3}$, which is well below the value 
of $10^6\,$cm$^{-3}$ assumed by Ossenkopf and Henning in their coagulation 
calculations. It is possible that the dense gas in the core of the filament 
is clumped.  Further studies of the correlation between 8$\mu$m extinction and 
submillimeter emission in this and other IRDCs should prove beneficial.

\clearpage

\begin{thebibliography}{}
\bibitem[Alves et al.\ (1998)]{all98} Alves, J., Lada, C.J., Lada,
E.A., Kenyon, S.J., \& Phelps, R. 1998, \apj, 506, 292
\bibitem[Becker et al.\ (1994)]{bec94} Becker, R.H., White, R.L.,
Helfand, D.J., \& Zoonematkermani, S. 1994, \apjs, 91, 347
\bibitem[Carey et al.\ (1998)]{car98} Carey, S.J., Clark, F.O., Egan,
M.P., Price, S.D., Shipman, R.F., \& Kuchar, T.A. 1998, \apj, 508, 721
\bibitem[Carey et al.\ (2000)]{car00} Carey, S.J., Feldman, P.A.,
Redman, R.O., Egan, M.P., MacLeod, J.M., \& Price, S.D. 2000, \apj, 543, L157
\bibitem[Economou et al.\ (1999)]{ebw99} Economou, F., Bridger, A.,
Wright, J.S., Jenness, T., Currie, M.J., Adamson, A. 1999, in
Astronomical Data Analysis Software and Systems VIII, 
ASP Conf.\ Ser.\ Vol.\ 172, eds. D.M. Mehringer, R.L. Plante, D.A. Roberts, 
(San Francisco: ASP), p.11
\bibitem[Egan et al.\ (1998)]{ega98} Egan, M.P., Shipman, R.F., Price,
S.D., Carey, S.J., Clark, F.O.,\& Cohen, M. 1998, \apj, 494, L199
\bibitem[Evans et al.\ (2001)]{ers01} Evans, N.II, Rawlings, J., Shirley, Y.,
\& Mundy, L. 2001, \apj, 557, 193
\bibitem[Fiege \& Pudritz (2000)]{fp00} Fiege, J.D. \& Pudritz, R.E. 2000,
\apj, 544, 830
\bibitem[Fiege et al.\ (2003)]{fj03} Fiege, J.D., Johnstone, D., Redman, R.O.,
Feldman, P.A., \& Carey, S. 2003, \apj, in preparation
\bibitem[Holland et al.\ (1999)]{hol99} Holland, W.S., et al. 1999, 
\mnras, 303, 659  
\bibitem[Jenness \& Economou (1999)]{je99} Jenness, T. \& Economou, F. 1999, in 
Astronomical Data Analysis Software and Systems VIII, 
ASP Conf.\ Ser.\ Vol.\ 172, eds. D.M. Mehringer, R.L. Plante, D.A. Roberts, 
(San Francisco: ASP) , p.171
\bibitem[Jenness et al.\ (2000)]{jen00} Jenness, T., Lightfoot, J.F.,
Holland, W.S., Greaves, J.S., \& Economou, F. (2000), in Imaging at
Radio through Submillimeter  Wavelengths,  eds. J. Mangum \& S. Radford
(San Francisco: ASP), p.205
\bibitem[Jenness et al.\ (2002)]{jsa02} Jenness, T., Stevens, J.A., Archibald,
E.N., Economou, F., Jessop, N.E., \& Robson, E.I. 2002, \mnras, 336, 14
\bibitem[Johnstone \& Bally (1999)]{jb99} Johnstone, D. \& Bally, J. 1999, 
\apj, 510, L49
\bibitem[Johnstone et al.\ (2000)]{jmi00} Johnstone, D., Wilson, C.D., 
Moriarty-Schieven, G., Creighton, J.G., \& Gregersen, E. 2000,
\apjs, 131, 505
\bibitem[Kramer et al.\ (2003)]{krm03} Kramer, C., Richter, J., Mookerjea, B.,
Alves, J., \& Lada, C. 2003, A\&A, 399, 1073
\bibitem[Lada et al.\ (1999)]{lal99} Lada, C.J., Alves, J., \& Lada, E.A. 1999,
\apj, 512, 250
\bibitem[Li \& Draine (2001)]{ld01} Li, A. \& Draine, B.T. 2001, \apj, 554, 778
\bibitem[Martin \& Whittet (1990)]{mar90} Martin, P.G. \& Whittet,
D.C.B. 1990, \apj, 357, 113
\bibitem[Li \& Lunine (2003)]{ll03} Li, A. \& Lunine, J.I. 2003, \apj, in press
\bibitem[Matthews et al.\ (2002)]{mfm02} Matthews, B.C., Fiege, J.D., \&
Moriarty-Schieven, G. 2002, \apj, 569, 304
\bibitem[Matthews \& Wilson (2000)]{mw00} Matthews, B.C. \& Wilson, C.D. 2000,
\apj, 531, 868
\bibitem[Matthews et al.\ (2001)]{mwf01} Matthews, B.C., Wilson, C.D. \& Fiege 
J.D. 2001, \apj, 562, 400
\bibitem[Nakamura et al.\ (1993)]{nhn93} Nakamura, F., Hanawa, T., \& 
Nakano, T. 1993, PASJ, 45, 551
\bibitem[Ossenkopf \& Henning (1994)]{oh94} Ossenkopf, V. \& Henning, Th.
1994, A\&A, 291, 943
\bibitem[Ostriker (1964)]{o64} Ostriker, J. 1964, \apj, 140, 1056
\bibitem[Redman et al.\ (2000)]{red00} Redman, R.O., Feldman, P.A.,
Carey, S.J., \& Egan, M.P. 2000, in 
Star Formation from the Small to the  Large Scale, 
Proc. 33$^{\rm rd}$ ESLAB Symp.\ ESA SP-445, eds.
F. Favata, A.A. Kaas, \& A. Wilson, p.499
\bibitem[Redman et al.\ (2003)]{red03} Redman, R.O., Feldman, P.A.,
Wyrowski, F., C\^{o}t\'{e}, S., Carey, S.J., \& Egan, M.P. 2003, ApJ, 586, 1127
\bibitem[Stepnik et al.\ (2003)]{sab03} Stepnik, B. et al.\ 2003, A\&A, 398, 551
\bibitem[Stod\'olkiewicz (1963)]{s63} Stod\'olkiewicz, J.S. 1963, Acta Astron.,
13, 30
\bibitem[Zucconi et al.\ (2001)]{zwg01} Zucconi, A., Walmsley, C.M., \&
Galli, D. 2001, A\&A, 376, 650
\end{thebibliography}
\end{document}